\documentclass[aps,epsfig,prl,twocolumn,superscriptaddress,showpacs,email,amssymb,nobibnotes]{revtex4}
\usepackage{graphicx,amsmath,amssymb}
\begin{document}
\title{Cluster approximations for infection dynamics on random networks}
\author{G. Rozhnova}
\author{A. Nunes}
\affiliation{Centro de F{\'\i}sica Te{\'o}rica e Computacional and
Departamento de F{\'\i}sica, Faculdade de Ci{\^e}ncias da Universidade de
Lisboa, P-1649-003 Lisboa Codex, Portugal}
\begin{abstract}
In this paper, we consider a simple stochastic epidemic model on large regular random graphs and the stochastic process that corresponds to this dynamics in the standard pair approximation. 
Using the fact that the nodes of a pair are unlikely to share neighbors, we derive the master equation for this process and obtain from the system size expansion the power spectrum of the fluctuations  in the quasi-stationary state. We show that whenever the pair approximation deterministic equations give an accurate description of the behavior of the system in the thermodynamic limit, the power spectrum of the fluctuations measured in long simulations 
is well approximated by the analytical power spectrum. If this assumption breaks down, then the cluster approximation must be carried out beyond the level of pairs. We construct an uncorrelated triplet approximation that captures the behavior of the system in a region of parameter space
where the pair approximation fails to give a good quantitative or even qualitative agreement. For these parameter values, the power spectrum of the fluctuations in finite systems can be computed analytically from the master equation of the corresponding stochastic process.
\end{abstract}
\maketitle
\section{I. \ \ Introduction}

Stochastic models on lattices are an old subject in statistical physics, and  
cluster mean-field theories have been developed and used both in the context of equilibrium
and non-equilibrium problems \cite{Ziman79,Marro99}. These are mean-field approximations that
are formulated in terms of the $n$-site joint probabilities. Since 
for any fixed $n$ the evolution equations for these probabilities do not in general form a closed set, the $(n+1)$-site probabilities are expressed in terms of the lower order probabilities
according to some closure assumption. At the lowest nontrivial order of truncation, this method corresponds to the well known pair approximation (PA) according to which the triplet probabilities $p(a,b,c)$ are factorized as $p(a,b,c)= p(a,b) p(c|b)$, where $a,b,c$ denote lattice nodes' states  and $p(c|b)$ is the conditional probability of having state $c$ in neighbor of a node in state $b$.
The PA is exact on the Bethe lattice or Cayley tree, a mathematical model which
cannot be realized as a physical system or simulated computationally. In finite lattices, 
in particular in the $d$-dimensional regular lattice with first neighbor interactions often
considered in these models, the PA is used  as the simplest
analytical description that includes an explicit representation of spatial correlations.
Despite being quantitatively inaccurate, it provides insight and qualitative information about
the system.  

More recently, the interest in this approximation procedure shifted from the exploration of
conceptual models to its applications in several problems of population dynamics.
The spread of a virus is an example of a dynamic process occurring on a discrete spatial arrangement that was modeled in the traditional literature with the 1-site or mean-field approximation (MFA) \cite{andersonmay,murray}. While the MFA reasonably reproduces the spreading behavior for topologies where the number of connections per node is either high or strongly fluctuating and for those that show small-world features, it is highly inaccurate for lattice 
and in general network structured populations.
The PA has become very popular in lattice-based stochastic 
models of ecological, epidemic and evolutionary game dynamics \cite{japoneses, tome1, sis, rand1, rauch, lebowitz, jerome, trapman, jorge, adaptive, shaw}, and several modifications of the PA
have been proposed for particular graphs and dynamic rules that lead to better quantitative agreement \cite{rand2, vanbaalen, filipe, petermann, szabo2,szabo1,ferguson,bauch}. Some of these modifications are based on different closure assumptions at the level of pairs, while others depend on including higher order clusters.

By contrast, the PA is expected to perform well, even quantitatively, for stochastic models on
large regular random graphs (RRGs) which are random networks of fixed connectivity per node, or degree \cite{hauert}. This is because of the RRG's statistical properties, namely the short loop density of the graph tending to zero in the limit of large graph size \cite{refgraphs1}, so that a RRG may be seen locally as a Bethe lattice of the same degree. 

In this paper, we consider the susceptible-infective-recovered-susceptible (SIRS) stochastic epidemic model on simple RRGs, that is RRGs with no loops formed by one or two edges.
We construct a detailed stochastic model based on the PA that captures the behavior of this dynamics in finite systems, in the sense that the power spectrum of the fluctuations computed analytically from the model matches the numerical power spectrum measured along the simulation runs whenever the averaged dynamics of the densities is well approximated by the solutions of the PA equations in the thermodynamic limit. This happens in a large region of parameter space because the quality of the approximation becomes poor only when the recovery rate is much larger than the rate of loss of immunity. The coupling between dynamics and graph structure intervenes in the microscopic description of the system through the different transitions that it is necessary to consider and the corresponding transition rates.

When recovery is much faster than loss of immunity and the PA fails in the thermodynamics limit, 
the cluster approximation must be carried out beyond the level of pairs. We show that a triplet approximation (TA) with a standard closure assumption is adequate to describe the behavior of the system up to recovery rates two orders of magnitude larger than the immunity waning rates. This parameter range is of interest in applications to childhood infectious diseases \cite{andersonmay, bauchearn}. 
In order to describe the behavior of finite systems in this region, a detailed stochastic model suitable for RRGs can be built on the basis of the TA following the procedure described for the PA.

\section{II. \ \ Deterministic and stochastic frameworks in the pair approximation}
In this section we propose a stochastic model of the SIRS epidemic process in the PA. The effects of the inclusion of an implicit representation of spatial dependence in a stochastic model have been recently studied \cite{ourpre}. An extension of the analysis of stochastic fluctuations from non-spatial models to the case of models on regular structures such as $d$-dimensional hypercubic lattices has also been performed \cite{mckane_lattice}. The present study elaborates on the stochastic models with implicit spatial dependence by including a detailed microscopic description of the transitions between the states of the nodes and the states of the pairs of the nearest neighbors, that can be treated analytically. The results obtained both for the averaged dynamics that describes the behavior of the system in the thermodynamic limit and for the spectrum of the fluctuations in the quasi-stationary state of finite systems agree well with the results of Monte Carlo simulations on RRGs in a large parameter range.
Technically, the PA approach is quite similar to that previously
developed for well-mixed systems also known as the MFA approach \cite{mckane_cycles} while having substantial differences which can be easily understood as soon as we review a few facts. 

The full set of deterministic equations describing the SIRS process in the MFA is written in terms of the densities of susceptible and infected populations (we will equivalently use the term "probability" in what follows). These differential equations are deduced on the assumption of uncorrelated nodes in the limit of infinite system size and constitute an approximation for the transient and quasi-stationary behaviors of large spatially extended systems. Accordingly, the associated stochastic model has two classes of individuals, infected and susceptible, as its independent dynamical variables, and a particular realization of the model at a given time $t$ consists of $m_1$ susceptible individuals, $m_2$ infected individuals and $(N-m_1-m_2)$ recovered individuals, where $N$ is the total population size or the number of nodes in a graph. 

The deterministic formulation of the SIRS model in the PA couples the dynamics of the node and pair densities in the thermodynamic limit. The purpose of the inclusion of pair densities is to improve the level of approximation in the description of spatially explicit simulations of the stochastic model. Consider the stochastic SIRS process on a RRG with fixed degree per node $k$ (RRG-$k$) and $N$ nodes. After the initial distribution of the nodes among the classes of susceptible ($S$), infected ($I$) and recovered ($R$) individuals, let the state of the system evolve in time according to asynchronous update of the events of infection, recovery and immunity waning. Namely, infected individuals recover at rate $\delta$ [$I\stackrel{\delta}{\rightarrow}R$], immunity of recovered individuals ceases at rate $\gamma$ [$R\stackrel{\gamma}{\rightarrow}S$], and infection in the susceptible-infected pairs of the nearest neighbors occurs with rate $\lambda$ [$SI\stackrel{\lambda}{\rightarrow}II$]. Each executed event taken separately changes the state of only one node, has its side-effects on the state of the pairs the node forms with its neighbors, propagates to larger configurations and finally to the whole graph. Consequently, frequency distributions or numbers of various correlations present in the graph, among which pairs are the simplest ones, change. To construct a stochastic model in the PA the transition probabilities of different events have to be approximated purely in terms of pairs.

More specifically, to represent independent classes of the constituents of the stochastic system we can choose five independent numbers subject to the constraints on the number of nodes and pairs in the graph. These can be five pair variables alone, one node and four pair variables or two node and three pair variables. We shall adopt the last variant so that taking the limit $N \to \infty$ in the stochastic PA model we recover the PA-SIRS deterministic equations in the form given in Ref. \cite{ourpre}.  
We introduce the following notation for the independent variables: $m_1$ and $m_2$ are the numbers of susceptible and of infected nodes and $m_3$, $m_4$ and $m_5$ are the numbers of susceptible-infected, susceptible-recovered and recovered-infected pairs of the nearest neighbor nodes, respectively, at time $t$. The full set of the five variables is denoted shortly as $m=(m_1,m_2,m_3,m_4,m_5)$. The remaining variables are not independent. They can be computed from the constraints as follows:
\begin{eqnarray}
\label{constraintsnodes} N &=&\sum\limits_{\alpha}{m_{\alpha}} \ , \\
\label{constraintspairs} km_{\alpha} &=& \sum\limits_{\alpha\neq\beta}{m_{\alpha\beta}}+2 m_{\alpha\alpha} \ , 
\end{eqnarray} 
\noindent{}where $m_{\alpha}$ is the number of nodes in state $\alpha$, $m_{\alpha\beta}=m_{\beta\alpha}$ is the number of pairs of the nearest neighbor nodes in states $\alpha, \beta \in \{S,I,R\}$. In this notation, $m_1$ and $m_2$ equal $m_S$ and $m_I$, and $m_3$, $m_4$ and $m_5$ equal  $m_{SI}$, $m_{SR}$ and $m_{RI}$, respectively. Note that in Eq. (\ref{constraintspairs}) the factor of 2 comes from the fact that $\alpha\alpha$ pairs must be counted twice. Further, the number of constituents in the classes, which are now interpreted as classes of individuals in specific states and classes of fixed contacts established amongst individuals in specific states, evolves according to the rule: whichever event is executed, a transition of one individual between the classes of individuals occurs simultaneously with transitions of the individual's $k$ contacts between the classes of contacts. This rule is nothing else than the statement of the local modifications on a RRG-$k$ caused by a single event: a change in state of one node induces the simultaneous change in states of the $k$ pairs whose common member is the node undergoing the transition. The coarse grained description where the effect of the change in state of a given node on the $k$ pairs that it forms is averaged over each pair type has been given in Ref. \cite{ourpre}. In this study, we consider a detailed stochastic model in which the $k$ pairs are, in general, different.
\begin{figure}
{\includegraphics[width=0.25\columnwidth]{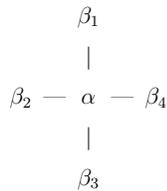} \caption{Schematic representation of a node in state $\alpha$ and its nearest neighbors in states $\beta_i$, where $i=1,\ldots,4$ and $\alpha, \beta_i\in\{S,I,R\}$, in a RRG-$4$.} \label{fig1}} 
\end{figure}

Next, we calculate the transition rates for the PA-SIRS stochastic process corresponding to the full microscopic description and yielding the PA deterministic equations as the equations of motion for the infinite system size. Consider a recovery event [$I\stackrel{\delta}{\rightarrow}R$] on a RRG-$4$ as an example. During a simulation of the stochastic process the rate of this event is equal to the constant rate of recovery, $\delta$, multiplied by the number of infected nodes at given time, $m_2$. As we keep track of the changes in states of the pairs formed by the central node that switches from infected to recovered, we have to calculate how this net rate is distributed among all possible configurations the infected node and its first neighbors might form. The configurations with a central node in state $I$ are accepted as different if they have different number of pairs $m_{I \alpha}$, where $\alpha \in \{S,I,R\}$, irrespective of their spatial arrangement. The total number of such configurations for recovery equals ${n+k-1 \choose k}$, where $n=3$ is the number of states and $k=4$ is coordination number of the graph. 

As it is well known in large sparse RRGs the members of a pair are unlikely to share neighbors \cite{refgraphs1,loops}. Having no nodes in common suggests that in the first approximation the pairs can be considered as independently distributed. In mathematical formulation it means that the probabilities of having particular configurations follow a multinomial law \cite{vanbaalen}. To be as general as possible we give the conditional probability of a configuration, shown in Fig. \ref{fig1}, by the formula: 
\begin{eqnarray}
\label{multinomial}
Q(\beta_1,\beta_2,\beta_3,\beta_4|\alpha) &\equiv & Q \left(\begin{smallmatrix}
   && \beta_1 &&\\
   &&|&&\\
   \beta_2&\text{---}&\alpha&\text{---}&\beta_4\\
   &&|&&\\
   &&\beta_3&&\\
   \end{smallmatrix}\right) = \nonumber\\
   &=& k! \prod_{i=1}^{k} \dfrac{Q(\beta_i|\alpha)^{m_{\alpha\beta_i}}}{m_{\alpha\beta_i}!} \ . 
\end{eqnarray}
\noindent{}Here $\alpha, \beta_i\in\{S,I,R\}$ denote states of the five nodes and $i=1,\ldots,k$. The number of neighbors to be independently distributed is $k=4$. $Q(\beta_i|\alpha)$ is the conditional probability that given a node in state $\alpha$ its nearest neighbor is in state $\beta_i$:
\begin{equation}
\label{conditionalinpairs}
Q(\beta_i|\alpha)=\left\{\begin{array}{l}\dfrac{m_{\alpha\beta_i}}{km_{\alpha}} \ \ \ \ \text{if} \ \ \alpha\neq\beta_i \ , \\
\dfrac{2m_{\alpha\beta_i}}{km_{\alpha}} \ \ \ \text{if} \ \ \alpha=\beta_i \ .
\end{array} \right.
\end{equation}
\noindent{}Summation over the states of all pairs present in a given configuration equals the number of the nearest neighbors: $\sum\limits_{\alpha\beta_i} m_{\alpha\beta_i} = k$.
\begin{figure}
{\includegraphics[width=0.68\columnwidth]{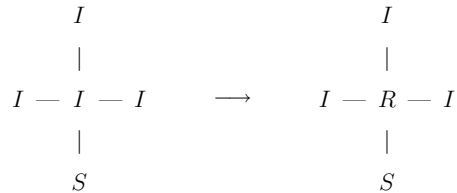}
\caption{Example of a configuration before and after recovery of the central node. The total number of recovery configurations is 15.}
\label{fig2}}
\end{figure}

If there are chains of connections between outer nodes of a configuration, especially short ones, the described method introduces an error because the pairs can no longer be considered independent. This is a crucial point that explains the failure of the standard PA for regular structures characterized by a huge number of loops of all lengths. For sparse RRGs with locally tree-like structure we assume the pairs to be uncorrelated, so that the probability distribution of a central node's neighbors can still be estimated by the multinomial Eq. (\ref{multinomial}). 

To demonstrate an application of Eq. (\ref{multinomial}) we choose a configuration before recovery of the central node shown in Fig. \ref{fig2}. The probability of this configuration is given by:
\begin{equation}
\label{multpart}
Q \left(\begin{smallmatrix}
   && I &&\\
   &&|&&\\
   I&\text{---}&I&\text{---}&I\\
   &&|&&\\
   &&S&&\\
   \end{smallmatrix}\right)=\dfrac{4!}{3!1!} Q(I|I)^3 Q(S|I)=\dfrac{4 (2m_{II})^3 m_{SI}}{(km_{I})^4} \ . \nonumber
\end{equation}
\noindent{}Using Eq. (\ref{constraintspairs}) and the notation introduced in the beginning of this section one obtains:
\begin{equation}
\label{multpartm}
Q \left(\begin{smallmatrix}
   && I &&\\
   &&|&&\\
   I&\text{---}&I&\text{---}&I\\
   &&|&&\\
   &&S&&\\
   \end{smallmatrix}\right) = \dfrac{4 (km_2-m_3-m_5)^3m_3}{(km_2)^4} \  .\nonumber
\end{equation}

According to Fig. \ref{fig1}, we introduce the total variation of the number of nodes in state $\alpha$ and of the number of pairs in state $\alpha\beta_i$ in a graph as the difference between the respective numbers after and before the central node switches from state $\alpha$ to another state: 
\begin{equation}
\Delta m_{\alpha} = m_{\alpha}^f-m_{\alpha}^i \ , \ \Delta m_{\alpha\beta_i} = m_{\alpha\beta_i}^f-m_{\alpha\beta_i}^i \ . \nonumber
\end{equation}
\noindent{}In Fig. \ref{fig2}, after recovery of the central node the configuration changes to that with a recovered node in the center inducing transitions in pairs of the nearest neighbors. The corresponding number variations for the independent variables are $\Delta m_1=0$, $\Delta m_2=-1$, $\Delta m_3=-1$, $\Delta m_4=1$, and $\Delta m_5=3$. Finally, we can give the transition rate for a recovery event in the center of the configuration shown in Fig. \ref{fig2} as follows:
\begin{equation}
\label{tr_rate}
\mathcal{T}^{m_1,m_2-1,m_3-1,m_4+1,m_5+3}_{m_1,m_2,m_3,m_4,m_5}=\delta m_2 \dfrac{4 (km_2-m_3-m_5)^3m_3}{(km_2)^4} \ , \nonumber
\end{equation}
\noindent{}where the subscript and the superscript of $\mathcal{T}$ denote the initial and the final states of the system, respectively.

\begin{figure}
{\includegraphics[width=0.7\columnwidth]{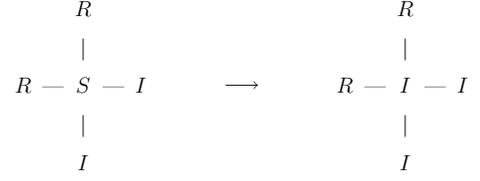}
\caption{Example of a configuration before and after infection. The total number of infection configurations is 10.}
\label{fig5}}
\end{figure}

The other transition rates for events of recovery and immunity waning are straightforward modifications of the above. As for infection process [$SI\stackrel{\lambda}{\rightarrow}II$] it is considered as inherent to a pair of nodes because transmission of infection requires a contact between two individuals, susceptible and infected. A transition of $S$ node in $SI$ pair from susceptible to infected changes both the state of this pair and the states of the other three pairs joining the $S$ node with its three nearest neighbors which we assume to be independently distributed in the PA. Thus, Eq. (\ref{multinomial}) is still applicable with $\alpha=S$, $\beta_4=I$ and $i=1,\ldots,k-1$:
\begin{eqnarray}
\label{simultinomial}
Q(\beta_1,\beta_2,\beta_3|SI) &\equiv& Q \left(\begin{smallmatrix}
   && \beta_1 &&\\
   &&|&&\\
   \beta_2&\text{---}&S&\text{---}&I\\
   &&|&&\\
   &&\beta_3&&\\
   \end{smallmatrix}\right) =\nonumber\\ 
   &=& (k-1)! \prod_{i=1}^{k-1} \dfrac{Q(\beta_i|S)^{m_{S\beta_i}}}{m_{S\beta_i}!} \ ,
\end{eqnarray}
\noindent{}where we have used the closure assumption in the standard PA, $Q(\beta_i|SI)\approx Q(\beta_i|S)$.
Now, for instance, the approximate transition rate for infection within an $SI$ pair of the configuration depicted in Fig. \ref{fig5} equals:
\begin{equation}
\label{sitr_rate}
\mathcal{T}^{m_1-1,m_2+1,m_3-2,m_4-2,m_5+2}_{m_1,m_2,m_3,m_4,m_5}=\lambda m_3\frac{3 m_3m_4^2}{(km_1)^3} \ .\nonumber
\end{equation}  

Having calculated all transition rates as described, we are ready to write the master equation for the PA-SIRS stochastic process \cite{vankampen,risken}:
\begin{equation}
\label{generalmasterequation}\dfrac{d\mathcal{P}(m,t)}{dt}=\sum\limits_{m'\neq m} \Big[\mathcal{T}^{m}_{m'}\mathcal{P}(m',t)-\mathcal{T}^{m'}_{m}\mathcal{P}(m,t) \Big] \ ,
\end{equation}
\noindent{}where $\mathcal{T}^m_{m'}$ denotes transition rates from other states $m'$ to state $m$ and vice versa for $\mathcal{T}^{m'}_{m}$. The complete solution of this differential-difference equation $\mathcal{P}(m,t)$ gives the probability of finding the system in state $m$ for all allowed sets of integers $m_i$, where $i=1,\ldots,5$, at time $t\geq0$ subject to the initial, normalization and boundary conditions. In general, it is not easy to solve this equation analytically but it is quite straightforward to analyze it for large but finite $N$ using van Kampen's system size expansion \cite{vankampen}. In that spirit, we set: 
\begin{equation}
\left\{ \begin{array}{l}
\label{nodefluc}
m_1(t)=NP(S)(t)+\sqrt{N}x_1(t) \ , \\ 
m_2(t)=NP(I)(t)+\sqrt{N}x_2(t) \ . 
\end{array} \right.
\end{equation}
\noindent{}In both equations, the first macroscopic terms scale with the system size $N$. The functions $P(S)(t)=\lim\limits_{N\to\infty}m_1(t)/N$ and $P(I)(t)=\lim\limits_{N\to\infty}m_2(t)/N$ are densities of susceptible and infected populations which have to be adjusted so as to satisfy the deterministic equations of motion in the PA. $x_1(t)$ and $x_2(t)$ are the new variables which denote stochastic fluctuations around the corresponding solutions of the PA deterministic equations and replace $m_1(t)$ and $m_2(t)$, respectively. The time-dependent transformations [Eq. (\ref{nodefluc})] from $m_1(t)$, $m_2(t)$ to $x_1(t)$, $x_2(t)$ involving functions $P(S)(t)$, $P(I)(t)$ come from the fact that one expects, with respect to the node variables, the probability distribution function $\mathcal{P}(m,t)$ to have a sharp peak around the macroscopic values $m_1(t)=NP(S)(t)$, $m_2(t)=NP(I)(t)$ with a width of order of $\sqrt{N}$, so that the functions $P(S)(t)$, $P(I)(t)$ follow the motion of the peak in time. Wheareas the system involves nodes and pairs as constituent elements we should carefully define what is meant by stochastic fluctuations around the solutions of the PA deterministic equations regarding the pair variables. For the fluctuations of the pair densities we set:
\begin{equation}
\left\{ \begin{array}{l}
\label{parefluc}
m_3(t)=NkP(SI)(t)+\sqrt{N}kx_3(t) \ , \\ 
m_4(t)=NkP(SR)(t)+\sqrt{N}kx_4(t) \ , \\ 
m_5(t)=NkP(RI)(t)+\sqrt{N}kx_5(t) \ .
\end{array} \right.
\end{equation}
\noindent{}In the above transformations, the first macroscopic terms are of order $Nk$, and the ansatz for the fluctuations scaling with the system size is $\sqrt{N}k$ for fixed parameter $k$ [actually, the scaling of the macroscopic terms in Eqs. (\ref{nodefluc}) and (\ref{parefluc}) can be found from Eqs. (\ref{constraintsnodes}) and (\ref{constraintspairs})]. The densities of susceptible-infected, susceptible-recovered and recovered-infected pairs are defined as $P(SI)(t)=\lim\limits_{N\to\infty}m_3(t)/(Nk)$, $P(SR)(t)=\lim\limits_{N\to\infty}m_4(t)/(Nk)$ and $P(RI)(t)=\lim\limits_{N\to\infty}m_5(t)/(Nk)$, respectively.  

The large-$N$ expansion is carried out using integer step operators $\epsilon_i$, where $i=1,\ldots,5$, that can be expressed as Taylor series involving partial derivatives with respect to the node and pair fluctuation variables \cite{vankampen}:
\begin{equation}
\label{epsilons}
\epsilon_{i}=\left\{ \begin{array}{l}
\sum\limits_{n=0}^{\infty}\dfrac{1}{n!}\left(\dfrac{1}{\sqrt{N}}\right)^n\dfrac{\partial^n}{\partial x_{i}^n} \ \ \ \ \text{if} \ \ i=1,2 \ , \\
\sum\limits_{n=0}^{\infty}\dfrac{1}{n!}\left(\dfrac{1}{\sqrt{N}k}\right)^n\dfrac{\partial^n}{\partial x_{i}^n} \ \ \ \text{if} \ \ i=3,4,5 \ .
\end{array} \right.
\end{equation}
\noindent{}With the aid of Eqs. (\ref{nodefluc})-(\ref{epsilons}), the master equation (\ref{generalmasterequation}) can be written in the following form: 
\begin{equation}
\label{master_epsilon}
\dfrac{d \Pi(x,t)}{dt}= \sum\limits_{m'\neq m} \left(\prod\limits_{i} \epsilon_i^{n_i} -1\right)\mathcal{T}^{m'}_{m}\Pi(x,t) \ ,
\end{equation}
\noindent{}where the probability distribution function $\Pi(x,t) \equiv \mathcal{P}(m,t)$ and $n_i$ are integer powers such that: 
\begin{equation}
\prod\limits_{i} \epsilon_i^{n_i} \mathcal{T}^{m'}_{m}\Pi(x,t)=\mathcal{T}^{m}_{m'}\Pi(x,t) \ . 	\nonumber
\end{equation}      
\noindent{}The leading terms in the power series expansion of Eq. (\ref{master_epsilon}) are of order $\sqrt{N}$. Collecting and setting them to zero yields a set of five first order differential equations. These are the PA-SIRS deterministic equations, see the Appendix for their explicit form. In next-to-leading-order, setting to zero the terms of order $N^0$, one gets a linear multivariate Fokker-Planck equation for the probability distribution function $\Pi(x,t)$ \cite{vankampen,risken}:

\begin{equation}
\label{mffokkerplanck}
\dfrac{\partial\Pi}{\partial t}=-\sum\limits_{i,j}{A_{ij}\dfrac{\partial (x_j\Pi)}{\partial x_i}}+\dfrac{1}{2}\sum\limits_{i,j}{B_{ij}\dfrac{{\partial}^2\Pi}{\partial x_i \partial x_j}} \ . 
\end{equation}

\noindent{}Here $x=(x_1,x_2,x_3,x_4,x_5)$ are stochastic fluctuations of the node and pair densities about their endemic steady state values in the PA. $\textbf{A}$ is the Jacobian matrix of the PA equations linearized about the endemic equilibrium solution, see formula (\ref{pajacobian}) in the Appendix. $\textbf{B}$ is symmetric internal noise cross correlation matrix derived directly from the expansion. For instance, $B_{12}=-\delta\bar{P}(I)$, where $\bar{P}(I)$ stands for the endemic equilibrium value of the density of infected individuals in the PA. The difference of the detailed stochastic description from the coarse grained description considered in Ref. \cite{ourpre} is reflected in the elements of the matrix $B_{ij}$, where $i,j=3,4,5$. 

Since the Fokker-Planck equation is linear, its solution $\Pi (x,t)$ is a multivariate Gaussian distribution completely determined by the first and the second moments. However, to analyze the fluctuations it is convenient to use the equivalent linear multivariate Langevin equation for $x_i(t)$ \cite{vankampen,risken}:  
   
\begin{equation}
\label{langevinequation}
\dot{x}_i(t)=\sum\limits_j{A_{ij}x_j(t)}+L_i(t) \ \ , \ \ i,j=1,\ldots,5 \ .
\end{equation}     
   
\noindent{}$L_i(t)$ are white random noise terms with the following properties:
\begin{equation}
\label{gaussian}
\left\{
          \begin{array}{ll}
          \left\langle L_i(t)\right\rangle =0  \ , \\
          \left\langle L_i(t)L_j(t')\right\rangle =B_{ij}\delta(t-t') \ .
          \end{array}
\right.
\end{equation}
\noindent{}The structure of the noise $x_i(t)$ as function of angular frequency $\omega$ is found from the power spectrum of the normalized fluctuations (PSNF) denoted as the averaged squared modulus of the Fourier transform of $x_i(t)$:
\begin{equation}
\label{ps}
\text{P}_i(\omega)\equiv\left\langle |\tilde x_i(\omega)|^2 \right\rangle \ ,
\end{equation} 
\noindent{}where
\begin{equation}
\tilde x_i(\omega)=\dfrac{1}{\sqrt{2\pi}}\int\limits_{-\infty}^{+\infty} x_i (t) e^{-i\omega t}dt \ .
\end{equation}
\noindent{}Solving for the Fourier transforms from the linear Eq. (\ref{langevinequation}) and using the correlation function in the frequency domain
$\langle \tilde L_i(\omega)\tilde L_j(\omega')\rangle=B_{ij}\delta(\omega +\omega')$, the approximate analytical expression for the PSNFs about the endemic equilibrium solution of the PA equations becomes:

\begin{equation}
\label{psnfgeneral}
\text{P}_i(\omega) = \sum\limits_{j,k}M_{ik}^{-1}(\omega)B_{kj}M_{ji}^{-1}(-\omega),
\end{equation}

\noindent{}where $M_{ij}(\omega)=\text{i} \omega \; \delta_{ij} - A_{ij}$. The PSNF of the susceptibles (of the infectives) is then obtained by setting $i=1$ ($i=2$, respectively). In this case, the PSNFs are of the form $p (\omega )/q(\omega )$, where $p(\omega )$ and $q(\omega )$ are polynomials in $\omega$ of order 8 and 10, respectively. Note that the same formula [Eq. (\ref{psnfgeneral})] is also valid for the PSNFs in the MFA taking $\textbf{A}$ as the Jacobian of the MFA-SIRS differential equations linearized about the mean-field endemic equilibrium solution and the noise cross correlation matrix $\textbf{B}$ computed directly from the expansion. 

\begin{figure}[h]
{\includegraphics[width=\columnwidth]{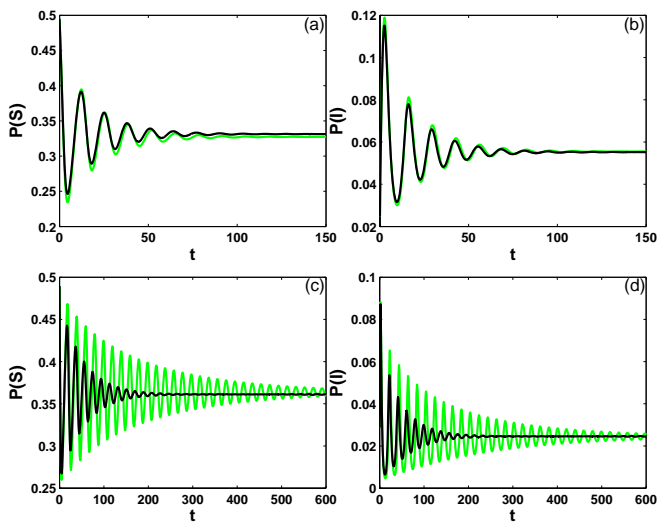}
\caption{(Color online) Densities of the infectives and of the susceptibles in the PA [gray (green) lines] and averaged numerical time series (black lines) obtained from Monte Carlo simulations of the SIRS stochastic model on a RRG-$4$ with $N=10^6$ nodes. All plots were obtained for $\delta=1$, $\lambda=2.5$ and $k=4$. Parameters: (a) and (b) $\gamma=0.09$; (c) and (d) $\gamma=0.04$.} 
\label{fig4}}
\end{figure}

\begin{figure}[h]
{\includegraphics[width=\columnwidth]{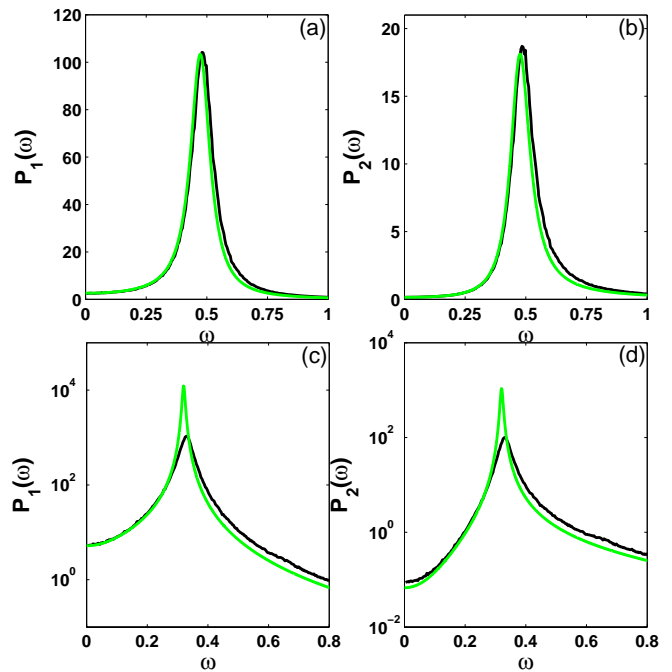}
\caption{(Color online) Analytical PSNFs of the infectives and of the susceptibles in the PA [gray (green) lines] and averaged numerical PSNFs (black lines) calculated from Monte Carlo simulations of the SIRS epidemic process on a RRG-$4$ with $N=10^6$ nodes. All plots were obtained for $\delta=1$, $\lambda=2.5$ and $k=4$. Parameters: (a) and (b) $\gamma=0.09$, lin-lin plot; (c) and (d) $\gamma=0.04$, lin-log plot.} 
\label{ps_rrg4_pa}}
\end{figure}

Figs. \ref{fig4} and \ref{ps_rrg4_pa} compare results of the theory developed so far with data of the SIRS stochastic model obtained from Monte Carlo simulations on a RRG-$4$. The RRGs are generated using a quick algorithm introduced in Ref. \cite{graphgen1}. The algorithm guarantees that for small degrees at least, which is the case here, the RRG-$k$ on $N$ nodes is generated uniformly at random, in the sense that all RRG-$k$ on $N$ nodes have asymptotically the same probability as $N\to\infty$. In the stochastic simulations, the system is set in a random initial condition with fixed node and pair densities, after which the states of the nodes are updated asynchronously according to the events of infection, recovery and loss of immunity. Figure \ref{fig4} shows averaged or global behavior of the densities as obtained from many runs of numerical simulations, namely for each set of given parameter values and initial conditions the simulations are averaged over $10^3$ realizations of a RRG. Figure \ref{ps_rrg4_pa} analyzes the fluctuations about the steady state densities observed in an individual run. More precisely, we compute the PSNFs in $1.5\times10^3$ parallel long simulation runs on RRGs numerically, and then we average. In both figures, the same parameter values in the endemic phase are used in the corresponding panels to ease a comparison of the results. Note that on a finite graph, the only true steady state of the SIRS process corresponds to all nodes being in susceptible state, that is why the steady states of the PA model are compared with the quasi-stationary states of large but finite systems. \newline

Plots in Fig. \ref{fig4} show susceptible and infective densities as function of time. Numerical solutions of the PA deterministic equations given in the Appendix are plotted in gray (green). Black lines are sample averages of the densities as obtained from $10^3$ realizations of a RRG-$4$ with $N=10^6$ nodes. For the SIRS model
the agreement of the solutions of the PA deterministic equations with the averaged dynamics on RRGs is sensitive to the rate of immunity waning $\gamma$, viz it deteriorates with decreasing $\gamma$ \cite{epjb}. The upper panels in Fig. \ref{fig4} are plotted for a set of parameter values for which the solutions of the PA equations reproduce the behavior of the averaged times series on a RRG-$4$ with a good accuracy both in the transient and in the quasi-stationary regime. The lower panels in Fig. \ref{fig4} plotted for a smaller value of $\gamma$ with the other parameters being fixed reflect the striking difference in the dynamics of the simulations (black lines) and of the PA deterministic model [gray (green) lines]. Although the steady state values of the densities are close, during the transient period the averaged time series attenuate more rapidly in the numerical simulations than the damped oscillations predicted by the PA. \newline  

Plots in Fig. \ref{ps_rrg4_pa} are the approximate analytical PSNFs [gray (green) lines] given by formula (\ref{psnfgeneral}) and the averaged numerical PSNFs (black lines) calculated from $1.5\times 10^3$ replicate simulation runs of the SIRS process on a RRG-$4$ with $N=10^6$ nodes. The upper panels in Fig. \ref{ps_rrg4_pa} plotted for parameter values used in the upper panels in Fig. \ref{fig4} show the agreement between the analytical and numerical PSNFs of the susceptibles and infectives. The analytical PSNFs [Eq. (\ref{psnfgeneral})] deduced from the detailed stochastic PA model with the transition rates calculated on the basis of Eqs. (\ref{multinomial}) and (\ref{simultinomial}) approximate the averaged numerical PSNFs well in those regions where the PA deterministic model predicts the same behavior as that of the times series obtained from the simulations averaged over many realizations. In this region, the typical PSNF is a bell-shaped curve indicating that in an individual simulation run susceptible and infective densities oscillate in time with frequencies close to the principal frequency demarked by the maximum of the curve. In fact, the large oscillations of the densities are due to a kind of internal resonance previously studied for a non-spatial predator-prey model \cite{mckane_cycles}. In the lower panels in Fig. \ref{ps_rrg4_pa} the amplitude of both the analytical and the numerical PSNFs gets much higher, indicating that for smaller values of $\gamma$ the stochastic fluctuations become more pronounced and better structured. The fact that the numerical PSNFs lie significally below those predicted analytically is closely related to the stability of the endemic equilibrium of the PA equations. As $\gamma$ decreases the equilibrium's stability gets weaker until it is lost on a critical line corresponding to a supercritical Andronov-Hopf bifurcation \cite{ourpre}. Accordingly, the approximate analytical PSNFs become more and more enhanced until they finally diverge on the critical line. Note that in the lower panels in Fig. \ref{fig4} plotted for the same parameters the behaviors of the analytical and numerical global densities do not agree too.        

A good correspondence and subsequent divergence between the analytical and numerical PSNFs can be understood from the van Kampen's large-$N$ expansion about the endemic equilibrium performed above. The analytical expression for the PSNFs, see formula (\ref{psnfgeneral}), is a function of $\omega$ and constant matrices $\textbf{A}$ and $\textbf{B}$ whose elementes are expressed in terms of the basic parameters of the model $\lambda$, $\gamma$, $\delta$ and $k$. In the general theory developed in Ref. \cite{vankampen}, the matrices can depend on time through the node [$P(S)$, $P(I)$] and pair [$P(SI)$, $P(SR)$, $P(RI)$] densities. However, as we are interested in the analysis of the fluctuations in the endemic equilibrium we have to substitute for the node and pair variables their stationary values [$\bar P (S)$, $\bar P (I)$, $\bar P (SI)$, $\bar P (SR)$, $\bar P (RI)$] depending on the parameters. This substitution results in time-independent coefficient matrices $\textbf{A}$ and $\textbf{B}$. It follows that the agreement between the analytical and numerical PSNFs is expectable in the regions where the PA deterministic model approximates the transient and quasi-stationary global dynamics on RRGs quite well. 
We emphasize that although the coarse grained stochastic model considered in Ref. \cite{ourpre} exhibits the same qualitative behavior as the detailed model described here, the quantitative agreement between the corresponding PSNFs is not achieved in that case. Therefore, the detailed microscopic description is needed to approximate the exact stochastic dynamics on RRGs.
Moreover, taking into account the above analysis it now becomes clear that resorting to higher order cluster approximations is necessary to describe the behavior of the SIRS stochastic model on RRGs for small $\gamma$. 

\section{III. \ \ Deterministic and stochastic frameworks beyond the pair approximation}

As mentioned in the previous section, in Ref. \cite{epjb} we have compared the data of the SIRS stochastic process obtained from Monte Carlo simulations on RRGs with the solutions of the standard PA equations and have shown that the PA describes correctly the global behavior of the model in the limit where rate of immunity waning $\gamma \gg 1$ but it fails to capture the dynamics for $\gamma \ll 1$. The question is then whether a cluster approximation of the next order can explain the suppression of global oscillations predicted by the PA for $\gamma \ll 1$ and, in particular, whether it can describe stationary states correctly. A reasonable description of the endemic equilibria as well as of the phase diagram of the stationary state calculated from numerical simulations is obtained by extending the generalized cluster approximation procedure to combinations of three neighboring nodes or triplets for short. 

In this section, we address this question and more generally the problem of the construction of cluster approximations of the order $q$ higher than two, that is higher than the PA. As a matter of fact, for the SIRS process the time evolution of the $q$-node joint probabilities is governed by a set of first order differential equations expressing their time derivatives as linear combinations of $q$- and $(q+1)$-node joint probabilities.
This is due to the infection process involving two nodes, susceptible and infected, simultaneously. In order to proceed the set of equations must be closed. In the standard cluster approximation the $(q+1)$-node joint probabilities are rational functions of the joint probabilities of smaller clusters of neighboring nodes, appropriately normalized, and thus the full set of first order differential equations can be obtained.

Within this perspective the PA is a standard cluster approximation for $q=2$. The TA is obtained for $q=3$ in a straightforward way by keeping node, pair and triplet probabilities as independent variables and expressing quadruplet probabilities in terms of them. We remind that each equation for the probability evolution of a particular cluster configuration is derived by considering all transitions leaving or entering it. Using the notation of the Appendix, in the TA of the SIRS process the equation for, for example, $P(RI)(t)$ reads as
\begin{equation}
\dfrac{dP(RI)}{dt}=\delta P(II)- (\delta+\gamma)P(RI)+\lambda (k-1)P(RSI) \ . \nonumber
\end{equation}
The first term on the right-hand side is due to the transition of $II$ pairs to $RI$ pairs occurring at rate $\delta$. The second [third] term corresponds to recovery [loss of immunity] of $I$ [$R$] node in $RI$ pairs that transit to $RR$ [$SI$] pairs at rate $\delta$ [$\gamma$]. Finally, a triplet $RSI$ changes to $RII$ with rate $\lambda$ such that a pair $RS$ is changed to $RI$. Since $S$ node of $RS$ pair has $(k-1)$ free neighbors that can infect it, this factor is included in the fourth term of the equation. Keeping again the same node [$P(S)$, $P(I)$] and pair probabilities [$P(SI)$, $P(SR)$, $P(RI)$] as independent variables the equation can be written in the following form: 
\begin{eqnarray}
\dfrac{dP(RI)}{dt}&=&\delta[P(I)-P(SI)]-(2\delta+\gamma)P(RI)+\nonumber\\
&+&\lambda (k-1)P(RSI) \ . \nonumber
\end{eqnarray} 
      
The equations for triplet probabilities are obtained in a similar way. In general, as far as clusters of more than two neighboring nodes are concerned, these split into two distinct classes, open and closed, and both have to be taken into account by considering the probabilities of finding an open and a closed configuration separately. By definition, an open cluster of a given size does not contain any loop while a closed cluster, on the contrary, necessarily contains at least one loop. Thus, for example, a triplet cluster can be closed forming a triangle or open forming a linear chain of three nodes. However, the number of short loops is small for large RRGs with small $k$ \cite{refgraphs1}. For instance, the analytical estimates for the mean $N_l$ and variance $Var(N_l)$ of the number of loops of length $l=3,4$ in a RRG-$4$ in the thermodynamic limit are $N_3=Var(N_3)=4.5$, $N_4=Var(N_4)=10.125$ \cite{loops}. The RRGs with $k=3$ are even more sparse, so that the number of short loops is even smaller. Based on these results we neglect the presence of small closed clusters in RRGs. Thus in the TA of the SIRS process, we complement Eq. (\ref{paeq}), see the Appendix, in which the probabilities of the triplets will be retained, with the equations for the triplet probabilities considering only open clusters of neighboring nodes. For instance, using the transition rules of the SIRS process the evolution equation for the probability $P(RRI)$ of finding a random triplet in state $RRI$ becomes:
\begin{eqnarray}
\dfrac{dP(RRI)}{dt}&=&\delta[P(RII)+P(IRI)-P(RRI)]-\nonumber\\
&-&2\gamma P(RRI)+\lambda(k-1)P(RRSI) \ ,\nonumber
\end{eqnarray}   
where $P(RRSI)$ is the probability of an open linear quadruplet in state $RRSI$. The deduction of other triplet equations is straightforward. 

To write the final closed set of differential equations in the TA we have: (a) to choose independent triplet variables whose number is reduced due to basic conservation relations; (b) to approximate quadruplet probabilities in terms of the node, pair and triplet ones. 

We discuss these two questions by order. With the use of probabilistic relations  
\begin{equation}
\label{tripletprobsum}
P(\alpha\beta)=\sum\limits_{\chi}P(\alpha\beta\chi)=\sum\limits_{\chi}P(\chi\alpha\beta)  
\end{equation}
and reflection symmetries
\begin{equation}
\label{tripletsym}
P(\alpha\beta\chi)=P(\chi\beta\alpha) \ ,
\end{equation}
where $\alpha,\beta,\chi \in \{S,I,R\}$, the number of independent triplet variables yields 9 and thus the total number of the TA equations is 14. Note that in finite RRGs the probability of triplets at given time equals:
 \begin{equation}
    		\label{pairprob}
				P(\alpha\beta\chi)=\left\{
	      	\begin{array}{ll}
        \dfrac{m_{\alpha\beta\chi}}{k(k-1)N} \ \ \ \text{if} \ \ \alpha\neq\chi \ , \\
        \dfrac{2m_{\alpha\beta\chi}}{k(k-1)N} \ \ \ \text{if} \ \ \alpha=\chi \ , \\
	      	\end{array}
				\right.
			\end{equation}
where we used the obvious notation for triplets. In this case, the factor of 2 is due to double counting of $\alpha\beta\alpha$ triplets. From Eqs. (\ref{tripletprobsum})-(\ref{pairprob}), the constraints on the number of pairs and triplets of the following type can be found [compare with Eqs. (\ref{constraintsnodes}) and (\ref{constraintspairs})]:
			\begin{equation}
			(k-1)m_{\alpha\beta}=\sum\limits_{\alpha\neq\chi}{m_{\alpha\beta\chi}}+2 m_{\alpha\beta\alpha} \ .
			\end{equation}
\begin{figure}
{\includegraphics[width=\columnwidth]{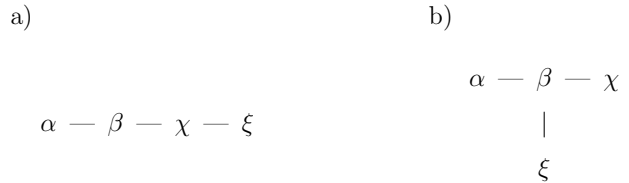}
\caption{(a) Linear and (b) "T-like" quadruplets both belong to the class of open quadruplet configurations, i.e. clusters of four nodes that do not contain any loop.}
\label{adds6}}
\end{figure}

A closer inspection of the equations for the triplets shows that probabilities of two types of open quadruplet configurations occur: linear quadruplets as in Fig. \ref{adds6} (a) and "T-like" quadruplets as in Fig. \ref{adds6} (b). Note that open triplet configurations can only be linear. With the assumption of uncorrelated triplets, linear quadruplets are approximated as follows. The joint probability of a quadruplet in state $\alpha\beta\chi\xi$ equals the joint probability to find a triplet in state $\alpha\beta\chi$ multiplied by the conditional probability to have a node in state $\xi$ neighboring with $\chi\beta\alpha$ triplet: 
\begin{equation}
\label{linquad}
P(\alpha\beta\chi\xi)=P(\alpha\beta\chi)Q(\xi|\chi\beta\alpha) \ .
\end{equation}  
Neglecting the influence of $\alpha$ node on $\xi$ node, the distribution of $\xi$ nodes in the neighborhood of $\chi\beta\alpha$ triplets is approximately equal to that in the neighborhood of $\chi\beta$ pairs:
\begin{equation}
\label{approx}
Q(\xi|\chi\beta\alpha)\approx Q(\xi|\chi\beta) \ . 
\end{equation}
Substitution of Eq. (\ref{approx}) into Eq. (\ref{linquad}) yields the closure assumption for linear quadruplets in the standard TA:
\begin{equation}
\label{clolin}
P(\alpha\beta\chi\xi)=\dfrac{P(\alpha\beta\chi)P(\beta\chi\xi)}{P(\beta\chi)} \ .
\end{equation}
Note the resemblance of Eq. (\ref{clolin}) with the closure assumption for open triplets in the standard PA:
\begin{equation}
P(\alpha\beta\chi)=\dfrac{P(\alpha\beta)P(\beta\chi)}{P(\beta)} \ .
\end{equation}
In both formulas the product of the probabilities (of pairs in the PA and triplets in the TA) is divided by the probability of the configuration in which they overlap (a node and a pair, respectively). Reasoning in the same way, a closure assumption for a "T-like" cluster depicted in Fig. \ref{adds6} (b), can be obtained. For example, considering a pair in state $\beta\xi$ as the overlapping configuration the probability a "T-like" quadruplet becomes:
\begin{equation}
\label{tastrong}
P\left(\begin{smallmatrix}\displaystyle
\alpha\beta\chi\\
\displaystyle\xi\\
\end{smallmatrix}\right)=\dfrac{P(\alpha\beta\xi)P(\xi\beta\chi)}{P(\beta\xi)} \ .
\end{equation}
Clearly, this closure assumption is not unique as any of the three pairs $\beta\alpha$, $\beta\chi$, and $\beta\xi$ joining in the node $\beta$ is suitable as an overlapping configuration. However, they all coincide in the limit when triplets are considered to be formed by uncorrelated pairs:
\begin{equation}
\label{taweak}
P\left(\begin{smallmatrix}\displaystyle
\alpha\beta\chi\\
\displaystyle\xi\\
\end{smallmatrix}\right)=\dfrac{P(\beta\alpha)P(\beta\chi)P(\beta\xi)}{P(\beta)^2} \ .
\end{equation}

Notwithstanding the closure assumption for "T-like" quadruplets, Eq. (\ref{tastrong}), is not unique it includes more information since the probability of a quadrupet is expessed in terms of the probabilities of triplets and pairs, unlike Eq. (\ref{taweak}) where the same probability depends on the probabilities of pairs and nodes. There is no a priori reason that the TA model with Eq. (\ref{tastrong}) should give a better approximation to Monte Carlo simulations than the same model with Eq. (\ref{taweak}) and direct numerical calculation of the distribution of nodes, pairs, triplets and quadruplets in the simulations is required to check this point. However, indirect evidence of this comes from the comparison of the TA model with both types of closure assumptions for "T-like" quadruplets and the stochastic SIRS process on RRGs in the regions where the PA model fails. We have found that the closure assumption given by Eq. (\ref{tastrong}) gives a quantitative improvement over the closure given by Eq. (\ref{taweak}) both for the stationary and for the time-dependent behaviors when it is faced with the exact stochastic dynamics on RRGs. Qualitatively, the behavior of both TA models is similar. For $k=4$ no stable oscillatory behavior is observed for small $\gamma$, and for $k=3$ the oscillatory phase becomes much smaller than in the PA model. The only difference we have been able to identify without performing the full linear analysis and calculating the whole phase diagrams of the TA deterministic models is that in the TA model with Eq. (\ref{tastrong}) the oscillations are suppressed faster than in the same model with Eq. (\ref{taweak}), making the former model a better approximation for the global behavior observed in the simulations on RRGs.

\begin{figure}[h]
{\includegraphics[width=\columnwidth]{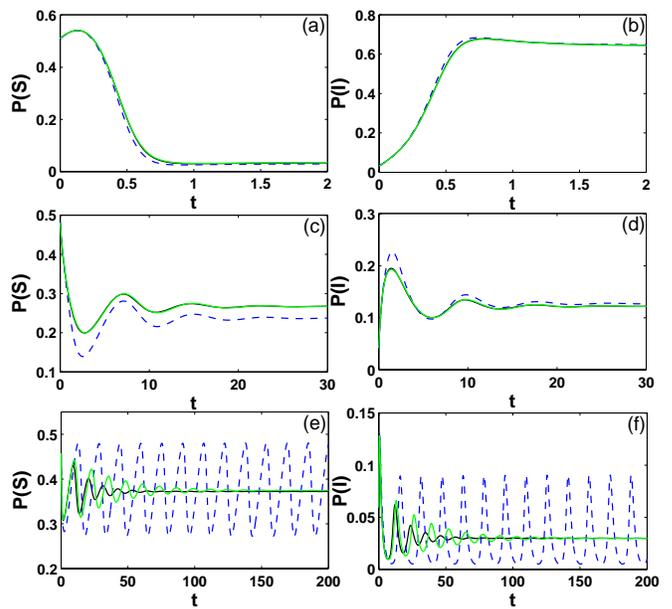}
\caption{(Color online) Time evolution of susceptible (left panels) and infective (right panels) densities for parameter values in the endemic region of the phase diagram of the PA model as predicted by the TA model [solid gray (green) lines], the PA model [dashed black (blue) lines] and the results of stochastic simulations (solid black lines) on a RRG-$3$ with $N=10^6$ nodes. All plots were obtained for $\delta=1$, $\lambda=15$ and $k=3$. Parameters: (a) and (b) $\gamma=2$; (c) and (d) $\gamma=0.2$; (e) and (f) $\gamma=0.05$.} 
\label{ta_pa_rrg3}}
\end{figure}

\begin{figure}[h]
{\includegraphics[width=\columnwidth]{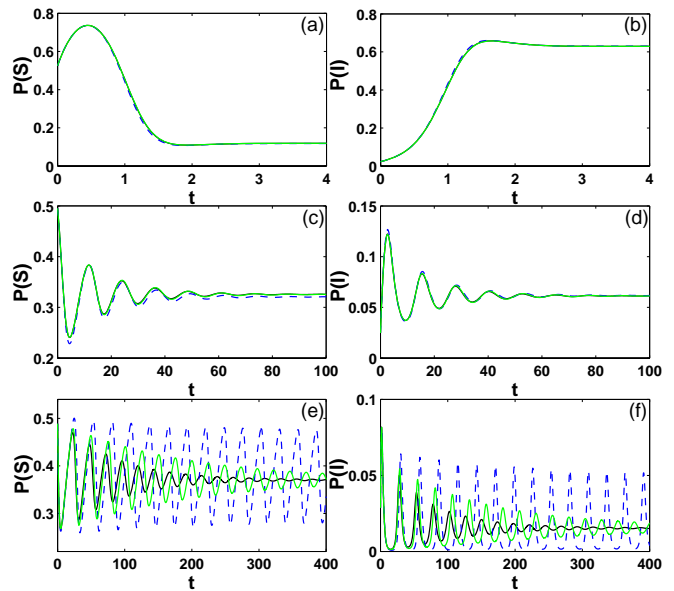}
\caption{(Color online) The same data as in Fig. \ref{ta_pa_rrg3} with $\delta=1$, $\lambda=2.5$ for $k=4$ and RRG-$4$ with $N=10^6$ nodes. Parameters: (a) and (b) $\gamma=2.5$; (c) and (d) $\gamma=0.1$; (e) and (f) $\gamma=0.025$.} 
\label{ta_pa_rrg4}}
\end{figure}

In Figs. \ref{ta_pa_rrg3} and \ref{ta_pa_rrg4} we compare the three models of the SIRS process, namely the TA-SIRS model with the closure assumptions for linear and "T-like" quadruplets given by Eqs. (\ref{clolin}) and (\ref{tastrong}) [solid gray (green) lines], the PA-SIRS model given in the Appendix [dashed black (blue) lines] and the results of stochastic simulations (solid black lines) on RRGs with $k=3$ and $k=4$, respectively. In the plots where the solid black lines cannot be distinguished they are superimposed by the solid gray (green) lines. The differential equations were integrated numerically using the 4th order Runge-Kutta algorithm, and for each set of initial conditions and parameters the simulations were averaged over an ensemble of $10^3$ RRGs of given degree with $N=10^6$ nodes. The properties of the fluctuations around this averaged behavior are described by the numerical PSNFs. For all parameter values considered below these numerical PSNFs are resonant-like as in Fig. \ref{ps_rrg4_pa}.  

In Fig. \ref{ta_pa_rrg3} the evolution of susceptible (left panels) and infective (right panels) densities as a function of time is shown for three sets of parameter values that correspond to constant infection rate $\lambda=15$ and decreasing rate of immunity waning $\gamma=2,0.2,0.05$ (from top to bottom). The values of the parameters are chosen so as to reflect the behavior of the models in three different phases in the endemic region of the phase diagram of the PA model \cite{epjb}. As regards the asymptotic behavior of the PA-SIRS equations, these phases are associated with the asymptotically stable nodes, asymptotically stable foci and limit cycles (from top to bottom). In the upper panels for the typical value $\gamma=2$ we used, the steady state of the system is a stable node both for the PA and for the TA model. The solutions of both deterministic models are almost coincident with the averaged densities obtained from the stochastic simulations for this whole region. The agreement between the deterministic models deteriorates in the region where both predict a behavior corresponding to a stable focus as can be seen in the middle panels where we used $\gamma=0.2$. The steady state of the PA equations is different from the quasi-stationary state observed in the simulations while the TA model describes the dynamics accurately both in the transient and in the steady state regime. The bottom panels with $\gamma=0.05$ illustrate that in the region where the PA model exhibits stable oscillatory behavior both the TA solutions and the simulation trajectories show damped oscillations towards a nontrivial equilibrium. The steady state densities given by the TA equations and the quasi-stationary densities calculated from the stochastic simulations on a RRG-$3$ are equal. However, in the transient regime we observe that the trajectories approach the steady state in a slightly different manner demonstrating a higher damping in the stochastic simulations. Also, for $k=3$ the oscillatory phase still persists in a very small region in the endemic phase in the TA but once more this result is not confirmed by the simulations for the same parameter values [results not shown]. Sustained oscillations, instead of resonant fluctuations, would show up as multiple peaks in the numerical PSNFs, which are not observed. This suggests that a complete description of the global behavior of the SIRS on a RRG-$3$ requires even more elaborate approximations and that the global oscillations predicted by both the PA-SIRS and the TA-SIRS in the thermodynamic limit are an artifact of the models. 

Fig. \ref{ta_pa_rrg4} illustrates the data as in Fig. \ref{ta_pa_rrg3} with $\lambda=2.5$, and $\gamma=2.5,0.1,0.025$ (from top to bottom) for $k=4$ and RRG-$4$. These values are chosen as before to represent the different phases of the PA diagram associated with the asymptotically stable nodes, asymptotically stable foci and limit cycles (from top to bottom). We observe the same comparative behavior of the models as $\gamma$ decreases. The only difference is that in the case of $k=4$ the oscillatory behavior in the TA has not been identified. Everywhere in the region of stable oscillations predicted by the PA model, the TA model reaches the steady state through damped oscillations that are associated with the existence of a stable fixed point, namely a stable focus. Such a dynamics of the TA-SIRS model is confirmed by the results of the simulations, see the bottom panels in Fig. \ref{ta_pa_rrg4}. 

Regarding the fluctuations, the same procedure that was used in the previous section to derive an approximate analytical expression for the PSNFs based on the PA can be extended to the TA in the region where the PA model fails. An explicit construction of a detailed stochastic TA model is cumbersome but straightforward, and once the master equation is written down the 
analytical PSNFs can be computed using van Kampen's system size expansion and then the general formula (\ref{psnfgeneral}). 
According to the results for the PA model, we expect the analytical PSNFs
obtained in this way to approximate well the numerical power spectra
in the parameter regions of the middle panels in Fig. \ref{ta_pa_rrg3}. We also expect a fair approximation in the parameter regions of the bottom panels of the same figure.

\section{IV. \ \ Discussion and conclusions}

The standard PA is known to perform poorly for lattice-based stochastic models but it gives in general good results for large RRGs due to the local tree-like structure of these graphs.
Taking a simple epidemic model as an example, we have used the PA to derive the master equation of the corresponding stochastic process on a RRG and an approximate analytical expression for the power spectrum of the fluctuations in the quasi-stationary state. 

We have checked the agreement of the analytical power spectrum in the PA against numerical simulations and found that whenever the behavior of the system in the thermodynamic limit is well described by the PA deterministic equations, the analytical power spectrum also describes accurately the fluctuations observed in long simulations.

This happens in a large region of parameter space. However, as $\gamma $ approaches the phase boundary $\gamma =0$ the quality of the PA deteriorates, and it is necessary to switch to higher order cluster approximations in order to obtain even qualitative agreement between the model equations and the simulations. We have shown that a TA with a standard closure assumption yields an accurate description of the behavior of the system in the thermodynamic limit in a $\gamma $ range where the PA breaks down. For finite systems, the fluctuation power spectrum 
can be computed analytically as before from the master equation of the stochastic process that corresponds to the TA. 

For small values of $\gamma $, long simulations require very large system sizes. For the smallest
$\gamma $ we have explored we found indications of the  breakdown of the TA, and that 
clusters of order higher than three would have to be considered. As the order of the cluster approximation increases, however, the 'no loop' assumption that is an ingredient of the construction of the detailed stochastic model becomes less accurate. Extension of this method to smaller values of $\gamma $ through higher order clusters would have to be combined with more complicated closure assumptions.

The parameter range explored in this paper is relevant to childhood infectious diseases modeling. Published estimates for the epidemiological parameters of measles, whooping cough, rubella and chicken pox, see Ref. \cite{andersonmay}, correspond in the SIRS model to $\gamma$ in the range $0.003 \sim 0.02$ assuming that the average immunity waning period can be taken as the typical duration of basic school. Our results show that the oscillatory phase that implicitly spatial models such as the PA-SIRS and the TA-SIRS exhibit in the thermodynamic limit cannot be directly related with the recurrent epidemic peaks found in many data sets \cite{andersonmay, bauchearn}. However, they also show that once stochastic effects are taken into account, the model predicts a well defined bell-shaped high amplitude fluctuation spectrum reproducing the qualitative features of typical time patterns of real data for this class of endemic diseases.

\textbf{\begin{center}
Acknowledgments
\end{center}}
Financial support from the Foundation of the University of Lisbon 
and the Portuguese Foundation for Science and 
Technology (FCT) under Contract No. POCTI/ISFL/2/618 is gratefully 
acknowledged. The first author (G.R.) was also supported by FCT under Grant No. SFRH/BD/32164/2006 and by Calouste Gulbenkian Foundation under its Program "Stimulus for  Research."

\section{Appendix}
Throughout the main text the node and pair probabilities are denoted as $P(\alpha)$ and $P(\alpha\beta)$, where the small Greek letters stand for states $S$, $I$, and $R$. We extend this notation for open linear triplet and quadruplet probabilities, $P(\alpha\beta\chi)$ and $P(\alpha\beta\chi\xi)$, respectively, while in the probabilities of open "T-like" quadruplets the clusters are depicted explicitly.      

Substituting $P(\alpha\beta\chi)=P(\alpha\beta)P(\beta\chi)/P(\beta)$ in the set below yields the PA-SIRS deterministic equations \cite{ourpre}: 
\begin{eqnarray}
\label{paeq}\dfrac{dP(S)}{dt}&=&\gamma[1-P(S)-P(I)]-k\lambda P(SI) \ , \\
\dfrac{dP(I)}{dt}&=&k\lambda P(SI)-\delta P(I) \ , \nonumber\\
\dfrac{dP(SI)}{dt}&=&\gamma P(RI)-\delta P(SI) -\lambda P(SI)+ \nonumber\\
&+&\lambda(k-1)[2P(SSI)+P(RSI)-P(SI)] \ , \nonumber\\
\dfrac{dP(SR)}{dt}&=&\delta P(SI)- \lambda(k-1) P(RSI)+\nonumber\\
&+&\gamma[1-P(S)-P(I)-P(RI)-2P(SR)] \ , \nonumber\\
\dfrac{dP(RI)}{dt}&=&\delta\left[P(I)-P(SI)-2P(RI)\right]-\gamma P(RI)+\nonumber\\
&+&\lambda (k-1) P(RSI) \ .\nonumber
\end{eqnarray}

The steady state solutions of the PA-SIRS equations can be obtained analytically. 
Let $\bar{P}(S)$, $\bar{P}(I)$ and $\bar{P}(SI)$, $\bar{P}(SR)$, $\bar{P}(RI)$ denote the endemic steady state values of the node and pair densities of the PA-SIRS model, then the Jacobian matrix $\textbf{A}$ of the linearized system is written as
\begin{widetext}
\begin{equation}
\label{pajacobian}
\textbf{A}=\left(
\begin{array}{ccccc} 
-\gamma & -\gamma & -k\lambda &  0  & 0                           \\
0  & -\delta & k\lambda & 0 & 0                      \\
C_0C_2 & 0 & C_3 & -\dfrac{C_1}{\bar{P}(SR)} & \gamma                      \\  
-\gamma+C_0 & -\gamma & \delta-\dfrac{C_1}{\bar{P}(SI)} & -2\gamma-\dfrac{C_1}{\bar{P}(SR)} & -\gamma \\ 
-C_0 & \delta & -\delta+\dfrac{C_1}{\bar{P}(SI)} & \dfrac{C_1}{\bar{P}(SR)} & -\gamma-2\delta
\end{array} \right) \ ,
\end{equation}
\end{widetext}  
where we introduced the constants   
\begin{equation}
C_0=\dfrac{(k-1)\lambda \bar{P}(SI)\bar{P}(SR)}{\bar{P}(S)^2} \ , \nonumber
\end{equation}
\begin{equation}
C_1=\dfrac{(k-1)\lambda \bar{P}(SI)\bar{P}(SR)}{\bar{P}(S)} \ , \nonumber
\end{equation}
\begin{equation}
C_2=1+2\frac{\bar{P}(SI)}{\bar{P}(SR)} \nonumber
\end{equation}
and
\begin{equation}
C_3=(k-2)\lambda-\delta-C_1\left(\dfrac{1}{\bar{P}(SI)}+\dfrac{4}{\bar{P}(SR)}\right) \ . \nonumber
\end{equation}
In the TA of the SIRS process, we complement Eq. (\ref{paeq}) in which the probabilities of the triplets are retained with the equations for the triplet probabilities as described in the main text.


\begin{thebibliography}{99}
\section{REFERENCES}

\bibitem{Ziman79}
J.~Ziman, \textit{Models of Disorder: The Theoretical Physics of Homogeneously Disordered Systems} (Cambridge University
Press, Cambridge, 1979). 

\bibitem{Marro99}
J.~Marro and R.~Dickman, \textit{Nonequilibrium Phase Transitions in Lattice Models} (Cambridge University Press, Cambridge, 1999).

\bibitem{andersonmay}
R.~M.~Anderson and R.~M.~May, \textit{Infectious Diseases of Humans: Dynamics and Control} (Oxford University Press, Oxford, 1991). 

\bibitem{murray}
J.~D.~Murray, \textit{Mathematical Biology I: An Introduction} (Springer-Verlag, New York, 2002).

\bibitem{japoneses}
H.~Matsuda, N.~Ogita, A.~Sasaki, and K.~Sato, Prog. Theor. Phys. {\bf 88}, 1035 (1992).

\bibitem{tome1}
J.~E.~Satulovsky and T.~Tom{\'e}, Phys. Rev. E {\bf 49}, 5073 (1994).

\bibitem{sis}
S.~Levin and R.~Durrett, Philos. Trans. R. Soc. London, Ser. B \textbf{351}, 1615 (1996). 

\bibitem{rand1}
D.~A.~Rand, in \textit{Advanced Ecological Theory: Principles and Applications}, edited by J.~McGlade (Blackwell Science, Oxford, 1999), p. 100.

\bibitem{rauch}
M.~A.~M.~de~Aguiar, E.~M.~Rauch, and Y.~Bar-Yam, Phys. Rev. E {\bf 67}, 047102 (2003).

\bibitem{lebowitz}
J.~Joo and J.~L.~Lebowitz, Phys. Rev. E \textbf{70}, 036114 (2004).

\bibitem{jerome}
J.~Benoit, A.~Nunes, and M.~M.~Telo~da~Gama, Eur. Phys. J. B \textbf{50}, 177 (2006).

\bibitem{trapman}
P.~Trapman, Math. Biosci. {\bf 210}, 464 (2007). 

\bibitem{jorge}
H.~Ohtsuki, M.~A.~Nowak, and J.~M.~Pacheco, Phys. Rev. Lett. {\bf 98}, 108106 (2007).

\bibitem{adaptive}
T.~Gross, Carlos~J.~Dommar~D'Lima, and B.~Blasius, Phys. Rev. Lett. {\bf 96}, 208701 (2006).

\bibitem{shaw}
L.~B.~Shaw and I.~B.~Schwartz, Phys. Rev. E {\bf 77}, 066101 (2008).

\bibitem{rand2}
M.~J.~Keeling, D.~A.~Rand, and A.~J.~Morris, Proc. R. Soc. London, Ser. B {\bf 264}, 1149 (1997).

\bibitem{vanbaalen}
M.~van~Baalen, in \textit{The Geometry of Ecological Interactions: Simplifying Spatial Complexity}, edited by U.~Dieckmann, R.~Law, and J.~A.~J.~Metz (Cambridge University Press, Cambridge, 2000), p. 359.

\bibitem{filipe}
J.~A.~N.~Filipe and M.~M.~Maule, Math. Biosci. {\bf 183}, 15 (2003).

\bibitem{petermann}
T.~Petermann and P.~De~Los~Rios, J. Theor. Biol. {\bf 229}, 1 (2004).

\bibitem{szabo2}
G.~Szab\'o, A.~Szolnoki, and R.~Izs\'ak, J. Phys. A {\bf 37}, 2599 (2004).

\bibitem{szabo1}
A.~Szolnoki and G.~Szab\'o, Phys. Rev. E {\bf 70}, 037102 (2004).

\bibitem{ferguson}
P.~E.~Parham and N.~M.~Ferguson, J. R. Soc., Interface {\bf 3}, 483 (2006).

\bibitem{bauch}
C.~T.~Bauch, Math. Biosci. {\bf 198}, 217 (2005).

\bibitem{hauert}
G.~Szab\'o and C.~Hauert, Phys. Rev. E {\bf 66}, 062903 (2002).

\bibitem{refgraphs1}
S.~Janson, T.~Luczak, and A.~Rucinski, \textit{Random Graphs} (Wiley, New York, 2000); E.~Marinari and R.~Monasson, J. Stat. Mech. P09004 (2004).

\bibitem{bauchearn}
C.~T.~Bauch and D.~J.~D.~Earn, Proc. R. Soc. London, Ser. B {\bf 270}, 1573 (2003).

\bibitem{ourpre}
G.~Rozhnova and A.~Nunes, Phys. Rev. E {\bf 79}, 041922 (2009).

\bibitem{mckane_lattice}
C.~A.~Lugo and A.~J.~McKane, Phys. Rev. E {\bf 78}, 051911 (2008).

\bibitem{mckane_cycles}
A.~J.~McKane and T.~J.~Newman, 
Phys. Rev. Lett. {\bf 94}, 218102 (2005).

\bibitem{loops}
The assumption on which the following deduction relies is the one used in the standard deterministic PA. Strictly speaking, it implies the complete absence of closed loops in a RRG-$k$ which is, of course, not realizable in practice. However, the classical results of graph theory show that in the thermodynamic limit the average expected number of short loops of length $l\ll \log N$, which are known to have the greatest effect on the performance of the PA, is given by $N_l=(k-1)^l/(2l)$ with Poisson fluctuations around the mean, so that it is negligibly small. These results are in a good agreement with the numerical values obtained from an ensemble of finite RRGs with $N$ ranging from $10^6$ to $5\times10^7$ nodes we generate in the simulations.

\bibitem{vankampen}
N.~G.~van~Kampen, \textit{Stochastic Processes in Physics and Chemistry} (Elsevier, Amsterdam, 1981).

\bibitem{risken}
H.~Risken, \textit{The Fokker-Planck Equation} (Springer, Berlin, 1996).

\bibitem{graphgen1}
A.~Steger and N.~C.~Wormald, Combinatorics, Probab. Comput. {\bf 8}, 377 (1999); J.~H.~Kim and V.~H.~Vu, \textit{STOC 2003: Proceedings of the 35th Annual ACM Symposium on Theory of Computing} (ACM, New York, 2003), p. 213.

\bibitem{epjb}
G.~Rozhnova and A.~Nunes, submitted to Eur. Phys. J. B (2009), e-print arXiv:0907.0335.

\end{thebibliography}
\end{document}